\documentstyle[eqsecnum,aps]{revtex}

\input{psbox.tex}

\begin{document}

\draft


\title{\bf Development of a tight-binding  potential for bcc-Zr. Application to the study of vibrational properties}
\author{Marcel Porta and Teresa Cast\'{a}n}

\address{Departament d'Estructura i Constituents de la Mat\`eria, Facultat 
de
F\'{\i}sica,\\ Universitat de Barcelona, Diagonal 647, 08028 Barcelona,
Catalonia, Spain.}


\maketitle

\begin{abstract}

We present a tight-binding potential based on the moment expansion of the
 density of states, which includes up to the fifth moment.
The potential is fitted to bcc and hcp Zr and it is applied to the computation
of vibrational properties of bcc-Zr. In particular, we compute the isothermal
elastic constants in the temperature range $1200K<T<2000$K
by means of standard Monte Carlo simulation techniques. The agreement with
experimental results is satisfactory, especially in the case of the stability of the
lattice with respect to the shear associated with  $C^\prime$.
 However, the temperature decrease of the Cauchy pressure is not reproduced.
The $T=0$K phonon frequencies of bcc-Zr are also computed. The potential
predicts several instabilities of the bcc structure, and a crossing of the
longitudinal and transverse modes in the (001) direction. This is in agreement
with recent ab initio calculations in Sc, Ti, Hf, and La.

\end{abstract} \vspace{0.8cm}


\pacs{PACS numbers:  62.20.Dc, 61.50.Lt}


\section{Introduction}

A fundamental problem in Condensed Matter Physics is how to project
microscopic interactions in many-body systems onto a description in terms
of a reasonably small number of degrees of freedom \cite{stoneham}.  
This is particularly important in computer simulation studies.  Atomic-level
simulations are, intrinsically, large-scale calculations and in spite of
the huge improvement nowadays available in computing capacity, an
efficient method to rapidly evaluate energies which also treat
forces in a physically realistic way, still remains a major
difficulty to be solved.

The computation of solid properties always requires a particular choice
for the interatomic potential.  In the choice there is a compromise between
physics and efficiency.  Computational efficiency
makes empirical or semiempirical potentials desirable, but at the
same time one should require that the underlying physics behind the model
potential is able to reproduce the properties of the system or at least
those of interest.

Several years ago Friedel \cite{friedel69} suggested that the starting
point in the description of transition metals (TM) is a band-picture with a
 strong
d-character.  In this sense, the remarkable parabola-like behaviour of the
cohesive energy and the bulk modulus exhibited by most TM as a function of
the number of d-electrons \cite{friedel69,ducastelle} clearly indicates
that cohesion is mainly dominated by the d-states.  This has motivated the
development of many-body potentials based on the tight-binding (TB)
approximation \cite{finnis84,rosato} (For a review see Ref. \cite{carlsson90}).
They are semiempirical in nature, which makes them very appealing for computer
simulation studies and at the same time they incorporate the band
character of the metallic cohesion so that the attractive part of the
interatomic energy turns out to be many body.

In the present work, we develop a TB potential based on the
 moment expansion of the density of states, which includes up to the fifth
 moment.  It has been suggested that this is the lowest order needed to
 reproduce the general trends in the elastic constants of hcp and fcc TM as a
 function of band filling \cite{nastar}.

 Zirconium, as is the case of other metals and
alloys, is close packed at low temperatures, but undergoes a structural phase
 transition of the Martensitic type to a bcc structure at higher temperatures
\cite{burgers,fisher,nishiyama}.  Some features of this phase
transformation in Zr were previously studied by using the second
 moment approximation TB model \cite{willaime89,willaime91}.  Here, we shall
 not focus on the structural phase transition itself but rather concentrate
 our interest on the elastic
properties of the high-temperature bcc-phase, which is accepted to be
mainly stabilized by entropy effects \cite{zener,friedel74,moroni}.
In particular we calculate the temperature behaviour of the relevant elastic
 constants by using standard Monte Carlo simulation techniques. The results thus
obtained are compared with the available experimental data. The agreement is
rather satisfactory, but it does not allow us to draw conclusions about the
 reliability of the potential.
This is provided by the computation of the phonon dispersion curves for the
bcc-Zr at $T = 0$K. Indeed, the comparison with recent ab initio calculations
\cite{persson} is indicative that most of the fundamental physics governing
 the vibrational
properties of bcc-Zr is contained in the interatomic potential model.

The paper is organized as follows.  The next section is
devoted to the construction of the interatomic potential.  In Sec.  III we
describe the fitting procedure to Zr.  In Sec.  IV we present and discuss
the results and in Sec.  V conclusions are drawn.


\section{Development of the interatomic potential}

The interatomic potential is developed following the tight binding
 bond model (TBBM) by A. P. Sutton et al. \cite{sutton,suttonllibre} 
 in the two-centre orthogonal approximation.
 The  basis set of the TB Hamiltonian only includes $d$ atomic orbitals and
 the crystal field interactions are neglected. 

   The cohesive energy is decomposed into two terms,

\begin{equation}
E_{coh} = E_{bond} + E_{pair}.
\label{ecoh}
\end{equation}
 The term $E_{pair}$ is an empirical pair potential which stands for
 electrostatic and exchange-correlation interactions \cite{sutton}, and
 the bond energy is

\begin{equation}
E_{bond} = \sum_{i} \int^{E_F}n_{i}(E)\left ( E - \epsilon_{i} \right ) dE,
\label{ebond}
\end{equation}
where $E_F$ is the Fermi energy, $\epsilon_{i}$ are the on-site
 Hamiltonian matrix elements in the atomic orbital representation,
and $n_{i}(E)$ are the local densities of states (LDOS).

 In this TB model, the s-d hybridization, which is known to make a considerable
 contribution to the cohesive energy of TM \cite{gelatt}, is not
treated appropriately. Nevertheless, it can be assumed that such contribution
is included implicitly in the pair term, as in the paper by  Girshick et al. \cite{girshick},
 or either that this contribution should be proportional to the d band width
and therefore it is included in the bond term.

The condition of local charge neutrality is fulfilled by defining local
 Fermi energies. This is done through the relation,
\begin{equation}
\int^{E_{F_i}} n_i(E) dE = N_d.
\end{equation}
 where $N_d$ is the number of $d$ electrons per site and the atomic orbital
 energy is chosen to be the energy zero, $\epsilon_i = 0 \, \, \forall i$.
 This method is equivalent to shifting the LDOS rigidly, whereas
 adjusting the on-site energies selfconsistently,
 as proposed by Sutton {\it et al.}, means that the shifts of the LDOS are  
 accompanied by a distortion of their shape.

 The LDOS are constructed from their moments, $\mu_i^{(n)}$, following the
 formalism of the recursion method of Haydock \cite{haydock} as if the moments
 corresponded to an $s$ band. In this way the computed LDOS are rotationally
 invariant \cite{inoue}.
 That is, from the second to the fifth moments we compute the coefficients
  $b_1$, $a_1$, $b_2$ and $a_2$ of the recursion method.
 Since  the coefficients $a_n$, $b_n$ are convergent oscillating series
 \cite{hodges} their limit $a_\infty$, $b_\infty$ is estimated as
 $a_{\infty} = (a_1 + a_2)/2$ and $b_{\infty} = (b_1 + b_2)/2$,
 and the coefficients $a_n$, $b_n$ with $n > 2$ are assumed to be equal to
 $a_\infty$, $b_\infty$ which gives rise to the well known square root
 terminator of the continued fraction expansion of the diagonal
 elements of the Green function \cite{hhk}.
  The integration of the LDOS in order to obtain the bond energy is performed
 numerically.

To completely specify the bond energy we need to define the functional form of
the off-diagonal Hamiltonian matrix elements in the atomic orbital
 representation, known as coupling strengths or hopping integrals.
 In the two-centre approximation the coupling strengths between sites $i$ and
 $j$ can be written as

\begin{equation}
\langle i\alpha | {\cal H} | j\beta\rangle =  \left ( 
\Phi_{\alpha\beta}^{\sigma} ({\bf r}_{ij}/r_{ij}) V_{dd\sigma} +
\Phi_{\alpha\beta}^{\pi} ({\bf r}_{ij}/r_{ij}) V_{dd\pi} +
\Phi_{\alpha\beta}^{\delta} ({\bf r}_{ij}/r_{ij}) V_{dd\delta} \right )
 R(r_{ij})
\end{equation}
where $\Phi_{\alpha\beta}({\bf r}_{ij}/r_{ij})$ is the angular dependence
  given by the symmetry
 of the $d$ orbitals, as shown by Slater and Koster \cite{slater},
 and $R(r_{ij})$ is the dependence on distance of the $dd\sigma$, $dd\pi$ and
 $dd\delta$ bonds, which is assumed to be equal for all of them.
The relative strength of the couplings $V_{dd\sigma}:V_{dd\pi}:V_{dd\delta}$
 are the canonical values -6 : 4 : -1.

For the radial dependence we take

\begin{equation}
R(r) = \left \{ 
\begin{array}{ll}
\exp(-qr), & r < r_1 \\
\exp(-qr)(a+br+cr^2)(r_2-r), & \, r_1 < r < r_2 \\
0 & r > r_2.
\end{array} \right .
\label{radial}
\end{equation}
where 
 $a$, $b$, and $c$ are constants fixed by the condition
that $R(r)$ and its two first derivatives must be continuous at
 $r_1$, which gives $a=(3r_1^2 - 3r_1r_2 +r_2^2)/(r_2 - r_1)^3$,
 $b=(-3 r_1 +r_2)/(r_2 - r_1)^3$ and $c=1/(r_2 - r_1)^3$.
 By construction, $R(r)$ is also continuous at the cutoff distance $r_2$.


Finally the pair term of the cohesive energy is taken to be

\begin{equation}
E_{pair} = \displaystyle \sum_{i, j(\ne i)} V_{pair}(r_{ij}),
\end{equation}
with
\begin{equation}
V_{pair}(r) = \left \{ \begin{array}{ll}
A(\exp(-p_1r) + \xi \exp(-p_2r)), & r < \tilde{r_1} \\
A(\exp(-p_1r) + \xi \exp(-p_2r))(\tilde{a} +\tilde{b}r +\tilde{c}r^2)(\tilde{r}_2 -r)^2, & \tilde{r}_1 < r < \tilde{r}_2 \\
0 & r > \tilde{r}_2.
\end{array} \right .
\end{equation}
where $\tilde{a}$, $\tilde{b}$, and $\tilde{c}$ are constants fixed 
 by the condition that $V_{pair}(r)$ and its first two derivatives must be
continuous at $\tilde{r}_1$, which gives
 $\tilde{a}=(6\tilde{r}_1^2 -4\tilde{r}_1 \tilde{r}_2 + \tilde{r}_2^2)/
 (\tilde{r}_2 - \tilde{r}_1)^4$, $\tilde{b}=2 ( -4 \tilde{r}_1 + \tilde{r}_2 )/
 (\tilde{r}_2 - \tilde{r}_1)^4$ and $\tilde{c}=3/(\tilde{r}_2 - \tilde{r}_1)^4$.
 By construction, $V_{pair}(r)$ and its first
 derivative is also continuous at $\tilde{r}_2$.


The model has, in principle, six fitting parameters: $A$, $\xi$, $p_1$, $p_2$,
 $V_{dd\delta}$ and $q$. Nevertheless, the number of electrons, $N_d$, and the
cutoff distances $r_1$, $r_2$, $\tilde{r}_1$, and $\tilde{r}_2$ are also used as
fitting parameters although they  are not completely free since they are
 constrained by their physical character.

\section{Fitting to Zr}

In this section we discuss the procedure used to fit the model
parameters to the zero temperature properties of Zr, both in the hcp and
 the unstable bcc phases.

Since the bcc  structure is only stable at $T > 1135$K, the values of some
 properties used in the fitting procedure
are extrapolated from the high temperature experimental data, whereas others
are obtained from ab initio calculations at $T = 0$K.
 Most of the ab initio results have been calculated in this work using the
 WIEN97 code
 \cite{wien}, which follows the Full Potential Linearized Augmented Plane Wave
 method within density functional theory. All the computed quantities are
obtained in the generalized gradient approximation (GGA) by Perdew, Burke and
Ernzerhof \cite{perdew} for the exchange-correlation potential.

The properties of bcc-Zr that we have considered in the fitting procedure are: the
cohesive energy, the lattice parameter, the elastic constants and the unrelaxed
 vacancy formation energy. Moreover, we have also taken some
 properties of hcp-Zr at $T = 0$K into account, mainly its elastic constants, lattice
 parameter, the $c/a$ ratio, and the energy difference between both structures.

\subsection{Fitting of the bond energy}

The coupling strengths are essentially short-range functions. Therefore, we
 impose that they fall to zero between the second and third nearest neighbours
 and choose  $r_2 = 1.24 a_{bcc}$.
For the matching point of the tail of the function we chose $r_1 = 0.92
 a_{bcc}$, which lies between the first and second nearest neighbours.
 Notice that in this way the tail
of the coupling strengths interferes with the fitting procedure.
The values chosen for these parameters are those which allow a better
reproduction of the properties of hcp-Zr.

An important property of the elastic constants exploited in the fitting
procedure is that the Cauchy pressure given by an interatomic potential
which satisfies the mechanical equilibrium conditions only depends on the
 many-body term of the potential, that is, the Cauchy pressure is independent of
 the pair term \cite{girshick}.
Considering both the bcc and the hcp structures we have three
Cauchy pressures to deal with.
One of these, the  Cauchy pressure  $C_{12} - C_{66}$ of the hcp lattice
is affected by an internal relaxation of the lattice under strain.
In order to avoid the determination of the internal relaxation and its
 effect on the Cauchy pressure during the fitting procedure, 
 as the  test value for this quantity, we
 use only its homogeneous contribution (without internal relaxation) obtained
 from the experimental Cauchy pressure by subtracting the inhomogeneous
 contribution determined from ab initio calculations.
 In the hcp lattice the elastic constants affected by internal relaxations are 
 $C_{11}$, $C_{12}$ and $C_{66}=(C_{11}-C_{12})/2$, and the inhomogeneous
 contribution to $C_{12}$ is the same as that of $C_{11}$ but with the
 opposite sign. Therefore, we need only to estimate the inhomogeneous
 contribution to $C_{11}$ which is
 $I \simeq 7 \, \, {\rm GPa}$ \cite{pviena}. Using the
 experimental values of the elastic constants together with the ab initio
 results, the three Cauchy pressures are,

\begin{equation}
\begin{array}{lcl}
  (C_{12} - C_{44})_{bcc} & = &  57.4 \, \, {\rm GPa} \mbox{\cite{pviena}} \nonumber \\
  (C_{12} - C_{66})_{hcp}^{homog.}  & = & 9 \, \, {\rm GPa}
 \mbox{\cite{fisher,pviena}}  \label{cp} \\
  (C_{13} - C_{44})_{hcp}  & = & 28 \, \, {\rm GPa} \mbox{\cite{fisher}}. \nonumber
 \end{array}
\end{equation}

Another property which strongly depends on the many-body term is the unrelaxed
 vacancy formation energy, $(E_V^f)^{\rm unrelx}$, since the contribution of
 the pair potential term to this quantity is equal to
its contribution to the cohesive energy.
For the present interatomic potential, in the bcc lattice we get the following
 approximate relation,

\begin{equation}
 (E_V^f)^{\rm unrelx} \simeq - E_{pair} - 0.52 E_{bond}. 
\end{equation}
Taking into account (\ref{ecoh}) we therefore get,
\begin{equation}
E_{bond} \simeq ((E_V^f)^{\rm unrelx} + E_{coh})/0.48 \simeq -8.0 \, {\rm eV}.
\label{valebond}
\end{equation}
 This result is consistent with the value predicted by the renormalized-atom
 method
of Gelatt et al. \cite{gelatt} provided that the contribution to the cohesive
 energy of the s-d hybridisation is assumed to be contained in
the bond term ($E_{ \rm d \, band \,  broadening} + E_{\rm s-d \,
  hybridization} \simeq -7.8 {\rm eV}$).

From the literature we obtain physical values for the number of
 electrons, $N_d = 2.536$ \cite{bassani}, band width, $W$=7.8 eV
 \cite{apettifor}, and the
 derivative of the band width with respect to the atomic volume,
 $-3 \Omega d\ln W/d\Omega = 3.97$
 \cite{apettifor,pettiforllibre}, which is related to the parameter $q$.
 In order to extract the value of the parameter $q$ from this relation 
  we approximate the band width as
 $W \propto \sqrt{\mu^{(2)}}$, that is, proportional to the square root
 of the second moment of the density of states \cite{ducastelle},
 and using the radial form
of the coupling strengths defined in Eq. (\ref{radial}), we obtain
 $q \simeq 4.36 a_{bcc}^{-1}$.
 In the present model the band width is $W = E_{top} - E_{bottom} =
 4 b_{\infty}$ which is proportional to the parameter $V_{dd\delta}$.
 Therefore we can use $W$ instead of $V_{dd\delta}$ as the fitting parameter
 without loss of generality.

 Using these physical values for the parametres $N_d$, $W$ and $q$, in the
 bcc lattice the interatomic potential gives a value of $E_{bond}
 \simeq -7.5$ eV, which is consistent
 with the value predicted by the unrelaxed vacancy formation energy
 (Eq. (\ref{valebond})). Nevertheless, the Cauchy pressure obtained with 
 these parameters, $(C_{12} - C_{44})_{bcc} \simeq 104 \, \,
 {\rm GPa}$, is unacceptably high (see Eq. (\ref{cp})).
 We have, therefore, to find an alternative way of fitting them.
 We shall proceed by paying more attention to the directly observable
 physical quantities such as cohesive energy, lattice parameter,
 elastic constants and vacancy formation energy rather than to the physical
 interpretation of the parameters. 

For a given value of $N_d$, the parameters $W$ and $q$ are determined
 numerically to reproduce the Cauchy pressure of the bcc lattice given in Eq.
 (\ref{cp}) and the bond energy, $E_{bond} = -8.0$eV (Eq. (\ref{valebond})).
We repeat this fitting procedure for different values of $N_d$ and compute the
Cauchy pressures of the hcp lattice. The results are given in Table \ref{tabI}.
These Cauchy pressures are computed using the lattice parameter and $c/a$
ratio of the hcp lattice at which this structure is stabilized by the
interatomic potential. For this purpose we use an arbitrary pair contribution
 fitted to the lattice parameter, cohesive energy and elastic constants of
 the bcc structure.
In all cases, $a_{hcp}$ and $c/a$ are close to the experimental and ideal values
respectively, although we observe a tendency of $c/a$ to increase as the
number of electrons is increased.
The value of $N_d$ which allows reproduction of the three Cauchy pressures
simultaneously (Eq. (\ref{cp})) is $N_d = 1.45$, which is substantially lower
 than its physical value, $N_d = 2.536$ \cite{bassani}. Nevertheless, there
 are  two additional reasons for adopting the value of $N_d = 1.45$, {\it i})
 TB studies on the relative stability of hcp and fcc lattices as a 
 function of $N_d$ \cite{bassani} have shown that the hcp lattice is
 only stable for $N_d < 2$ (in the range $0.5 < N_d < 4.5$), and
 {\it ii}) the low value of the $c/a$ ratio in group IV hcp metals can be justified
in terms of the different contribution to the bond energy from nearest
neighbours located on the basal plane and those located off it  
 only if $N_d < 2$ \cite{sutton,finnis88}.

\begin{figure}
 \psboxto(0.99\textwidth;0cm){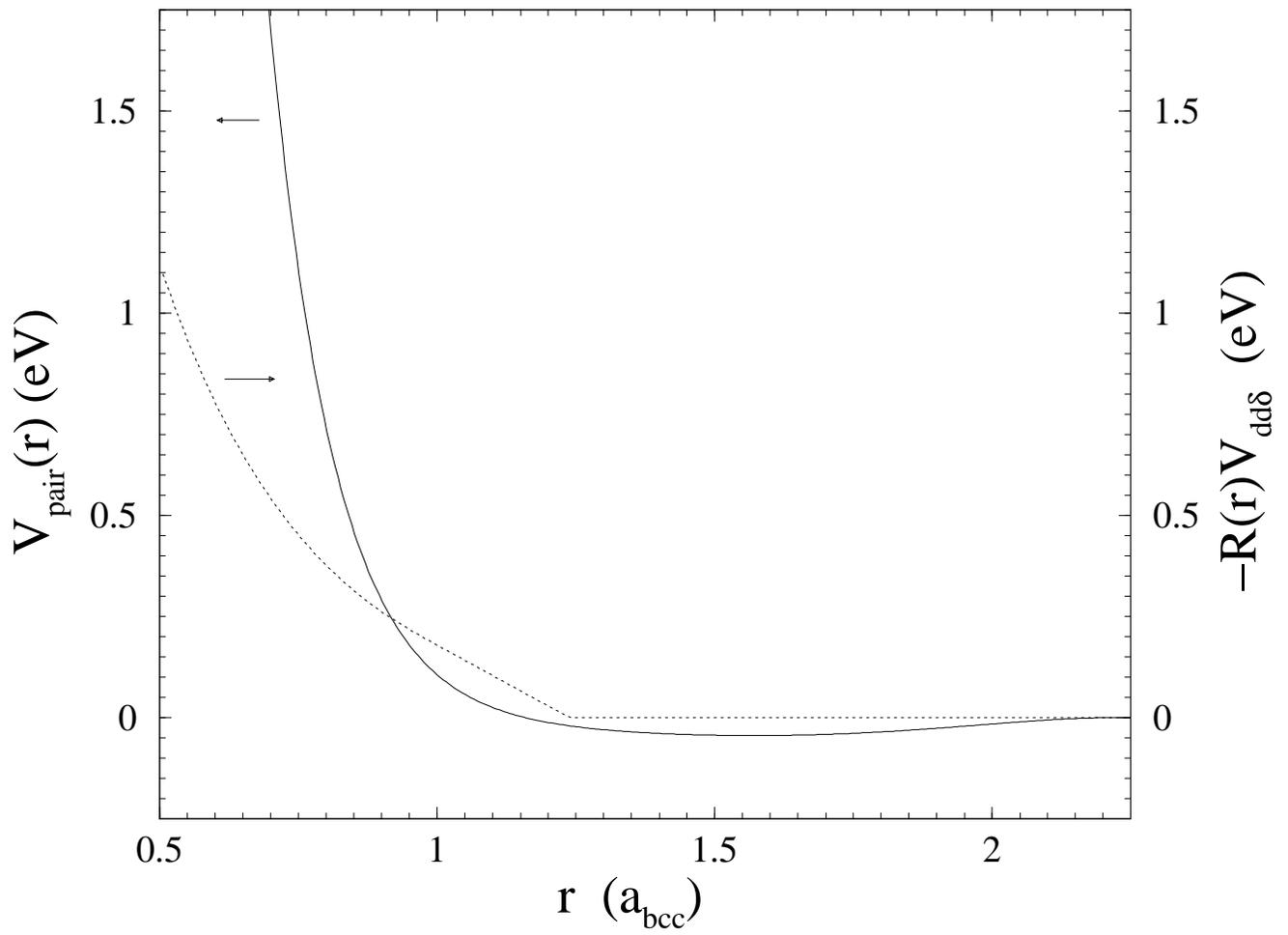}
\caption{Pair potential (solid line) and radial term of the coupling strengths
 (dashed line) {\sl vs.} interatomic distance.}
\label{curves}
\end{figure}

The radial dependence of the coupling strengths is shown in Fig. \ref{curves}.

\subsection{Fitting of the pair potential}

The cutoff distance of the pair potential is chosen to be $\tilde{r}_2 =
 2.211 a_{bcc}$ which lies between the 7th and  8th nearest neighbours in the
 bcc structure.
 This value has been adjusted in order to reproduce the properties of
 the hcp lattice.
 The value of $\tilde{r}_1$ is almost irrelevant since it does not really affect
 the functional form of the pair potential. We chose $\tilde{r}_1 = 1.2
 a_{bcc}$.
 For given values of the parameters
 $p_1$, $p_2$ and $\xi$, the parameter $A$ is determined from the mechanical
 equilibrium condition of the bcc lattice

\begin{equation}
\left . \frac{\partial E_{coh}}{\partial a} \right |_{a = a_{bcc}} = 0.
\end{equation}

The parameters $p_1$, $p_2$ and $\xi$ are given by the cohesive energy, and 
elastic constants $C_{11}$ and $C_{12}$ of the bcc lattice.   
Notice that the elastic constant $C_{44}$ is  automatically reproduced by
 fitting $C_{12}$ since the Cauchy pressure is fixed by the bond energy.

The functional form of the pair potential is shown in Fig. \ref{curves}.

 The values of the fitted parameters are given in Table \ref{tabII}
 and the properties of bcc and hcp Zr used for the fit are given in Tables
 \ref{tabIII} and \ref{tabIV} respectively.
 In the fitting procedure we have priorized the reproduction of the properties
 of bcc-Zr since the interatomic potential will be used to study the
 vibrational properties of this structure. This have led to different
 accuracy in the fitted quantities of bcc and hcp Zr (see Tables \ref{tabIII}
  and \ref{tabIV}).

\section{Calculation of the vibrational properties of bcc-Zr}


In this section we use the interatomic potential already obtained to compute
 the phonon dispersion curves of the bcc lattice at $T=0$K,
 and the elastic constants in the temperature range $1200{\rm K} < T < 2000$K, where 
experimentally, the bcc structure is stable.

\subsection{Phonon frequencies of bcc-Zr at zero temperature}

In order to obtain the phonon frequencies of bcc-Zr, we first compute the
 interatomic force constants by means of standard numerical differentiation.
 It is worth pointing out
 that the interatomic potential gives rise to long-range force constants. In
particular, the range of the pair potential contribution is the range of the
potential, which in the present case is  up to the 7th nearest neighbours. The
 range of the bond term contribution is much larger. For coupling strengths
 extending up to the second nearest neighbours and including up to the fifth
 moment
of the LDOS in the computation of the bond energy, the range of the force
 constants extends up to the 22th nearest neighbours. This is due to the
many-body character of the bond energy together with the dependence of the high
 order moments on the position of distant atoms.

\begin{figure}
 \psboxto(0.99\textwidth;0cm){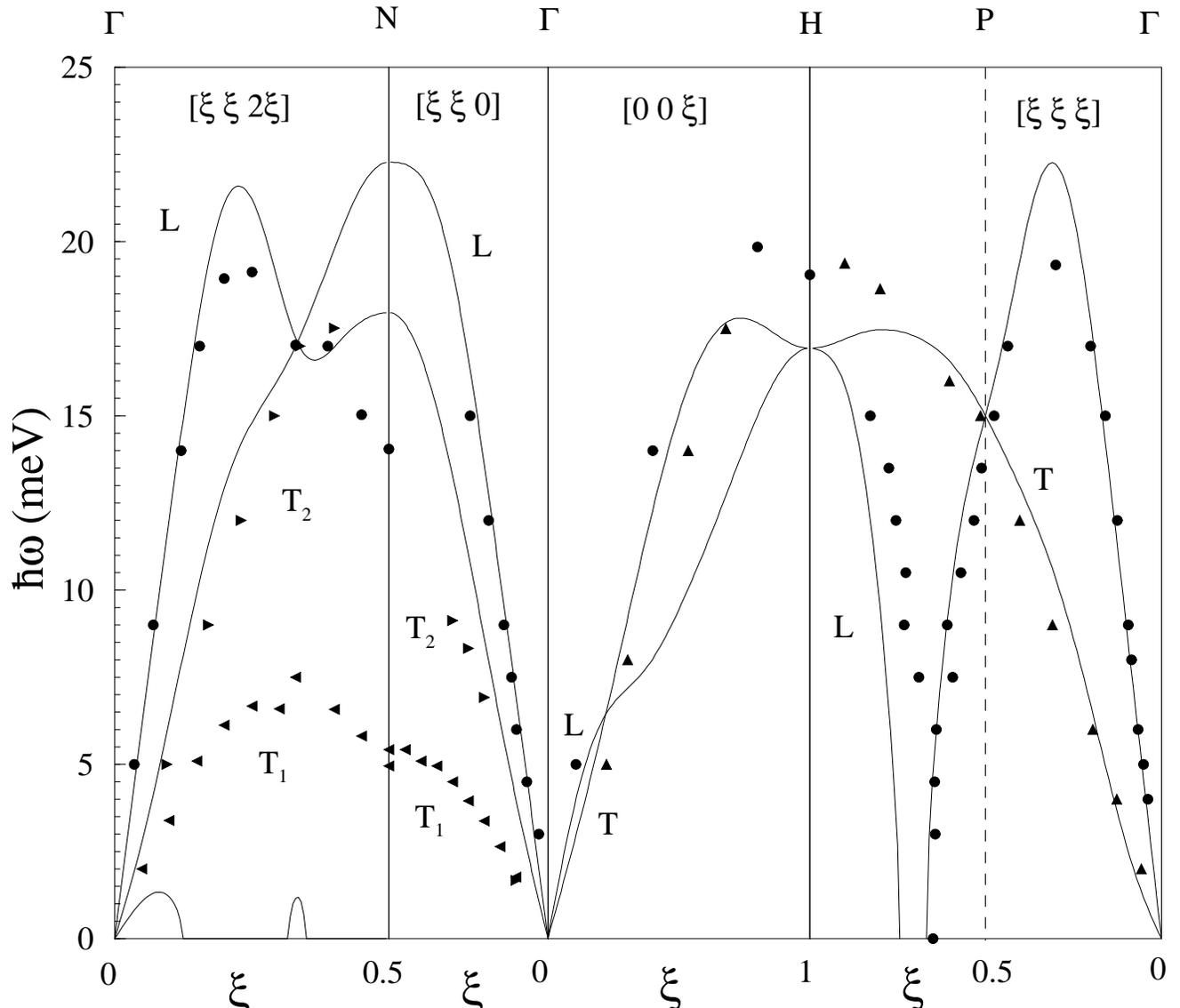}
 \caption{Computed phonon dispersion curves for bcc-Zr at T = 0 K
 (solid line) and experimental results  at 1200-1500K from
 Heiming et al. [37] (symbols).}
\label{fphonons}
 \end{figure}

In Table \ref{tabV} we show the results for the computed force constants.
 From these values we construct the dynamical matrix and compute the phonon
 frequencies in the harmonic approximation along the high symmetry directions of
 the reciprocal space (Fig. \ref{fphonons}). 
Since the interatomic potential is fitted to the $T = 0$K 
 elastic constants of bcc-Zr, the slope of the phonon dispersion curves around
 the $\Gamma$ point is expected to be correct.
 There are several features of the phonon dispersion curves that deserve
 special comment:

\begin{enumerate}

\item The whole T1($\xi$ $\xi$ 0) branch is unstable.

\item   
The T1($\xi$ $\xi$ $2\xi$) branch has a positive slope around the $\Gamma$
point (consistent with the associated combination of elastic constants), but it
 rapidly softens and becomes unstable. Before matching the T1($\xi$ $\xi$ 0)
 branch at the $N$ point, it becomes stable again at about $\xi = 1/3$. 

\item The softening of the L($\xi$ $\xi$ $\xi$) branch around
 $\xi = 2/3$  observed experimentally
 at high temperature is, at zero temperature, an instability which gives rise to the
$\omega$ phase. The frequency of the $\xi=2/3$ phonon is approximately zero,
as was obtained by K.M. Ho et al. \cite{ho} from ab initio calculations,
and the minimum of the branch is at about $\xi = 17/24$

\item In the (001) direction there is a crossing between the longitudinal and
transverse branches.

\end{enumerate}

 Several of these features, not observed experimentally at high temperature,
 were obtained from ab initio calculations in Sc, Ti, Hf, and La at $T=0$K
 \cite{persson}.
 In these materials the whole T1($\xi$ $\xi$ 0) branch has imaginary frequencies
(the phonons in the (112) direction have not been computed). The instability
around the $\xi=2/3$ L($\xi$ $\xi$ $\xi$) mode is also predicted, although the
minimum of the branch is not located at $\xi =17/24$ but at
$\xi = 7/12$ for all elements (except Sc, which has the minimum at $\xi = 2/3$).
Finally, the crossing of the longitudinal and transverse branches in the (100)
 direction is also observed.

 Moreover, in these
materials the T($\xi$ $\xi$ $\xi$) branch has imaginary frequencies around the $\Gamma$
point. The slope of this branch is given  by the elastic constant
 $C_{11} - C_{12} + C_{44}$, which in Zr is positive, and thus this feature
 cannot be observed.

From the force constants we also compute the elastic constants following the
 method of long waves \cite{bornhuang}, and recover the values obtained by means
of homogeneous deformations. This can be used  as a test of the internal
consistency of the interatomic potential \cite{sutton}.

\subsection{High temperature elastic constants of bcc-Zr}

The bcc-Zr high-temperature elastic constants are obtained from Monte Carlo
(MC) simulations in the canonical ensemble, $(T, V, N)$, using the atomic
 volume  obtained from MC simulations in the isobaric-isothermal
 ensemble, $(T, P, N)$, at zero pressure.
 The theoretical background of such simulations can be found in the work by
 I. R. McDonald \cite{mcdonald}.

The second-order isothermal elastic constants are computed using the
fluctuation formula \cite{lutsko}

\begin{eqnarray}
\begin{array}{l} \vspace{0.5cm}
\displaystyle   C_{ijkl}^T = \frac{1}{V} \frac{\partial^2 {F}}{\partial \epsilon_{ij} \partial \epsilon_{kl}} = 
  \frac{1}{V} \left \langle \frac{\partial^2 E_{coh}}{\partial \epsilon_{ij}
 \partial  \epsilon_{kl}} \right \rangle - \\
 \vspace{0.4cm} 
 \displaystyle  - \frac{1}{k_BTV} \left \{ \left \langle \frac{\partial E_{coh}}
  {\partial \epsilon_{ij}} \frac{\partial E_{coh}} {\partial \epsilon_{kl}}
  \right \rangle -
  \left \langle \frac{\partial E_{coh}} {\partial \epsilon_{ij}} \right \rangle
  \left \langle \frac{\partial E_{coh}} {\partial \epsilon_{kl}} \right \rangle 
   \right \} +    
 \displaystyle  \frac{Nk_BT}{V} (\delta_{il} \delta_{jk} + \delta_{ik} \delta_{jl}), 
\end{array}
\label{fluctuation}
\end{eqnarray}
where ${F}$ is the Helmholtz free energy, $\epsilon_{ij}$ are the
elastic strains, $V$ is the total volume,
  $k_B$ is the Boltzmann constant, $T$ is the temperature,
 and $N$ is the number of atoms.

The derivatives of the cohesive energy with respect to the elastic strains are
 computed numerically. Since the second derivative of both the pair potential
and the coupling strengths is not continuous at the cutoff distance, the
numerical derivative must be calculated using the same radial and pair functions in
 both the strained and the unstrained state of the lattice, for each of
 the atomic pairs. That is, if for a given
pair of atoms the distance, $r$,  is $r_1 < r < r_2$, the radial function used
 for the
computation of the bond energy of both the strained and unstrained lattice
will be that defined in this region, regardless of the fact that in the
 strained lattice we may have $r > r_2$.

The simulations are performed on a $4 \times 4 \times 4 \times 2 =
128$-site bcc lattice with periodic boundary conditions.
 The attempted configurational changes are single atom displacements and after
 each $N$ of these attempts (= 1 Monte Carlo step (MCS)), in the
 isobaric-isothermal  ensemble a volume change is also proposed.
The simulations are $10^5$ and $5\cdot10^4$ MCS long in the isobaric-isothermal
 and canonical ensemble respectively.

 Due to the many-body character of the bond energy, the change in the total
 cohesive energy due to a single-atom movement involves the recalculation of
the contribution to bond energy of about 65-110 atoms, depending on 
temperature. Nevertheless, this computation can be highly optimized and
only requires about six times the CPU  time needed to compute the contribution
to the bond energy of a single atom.

\begin{figure}
 \psboxto(0.99\textwidth;0cm){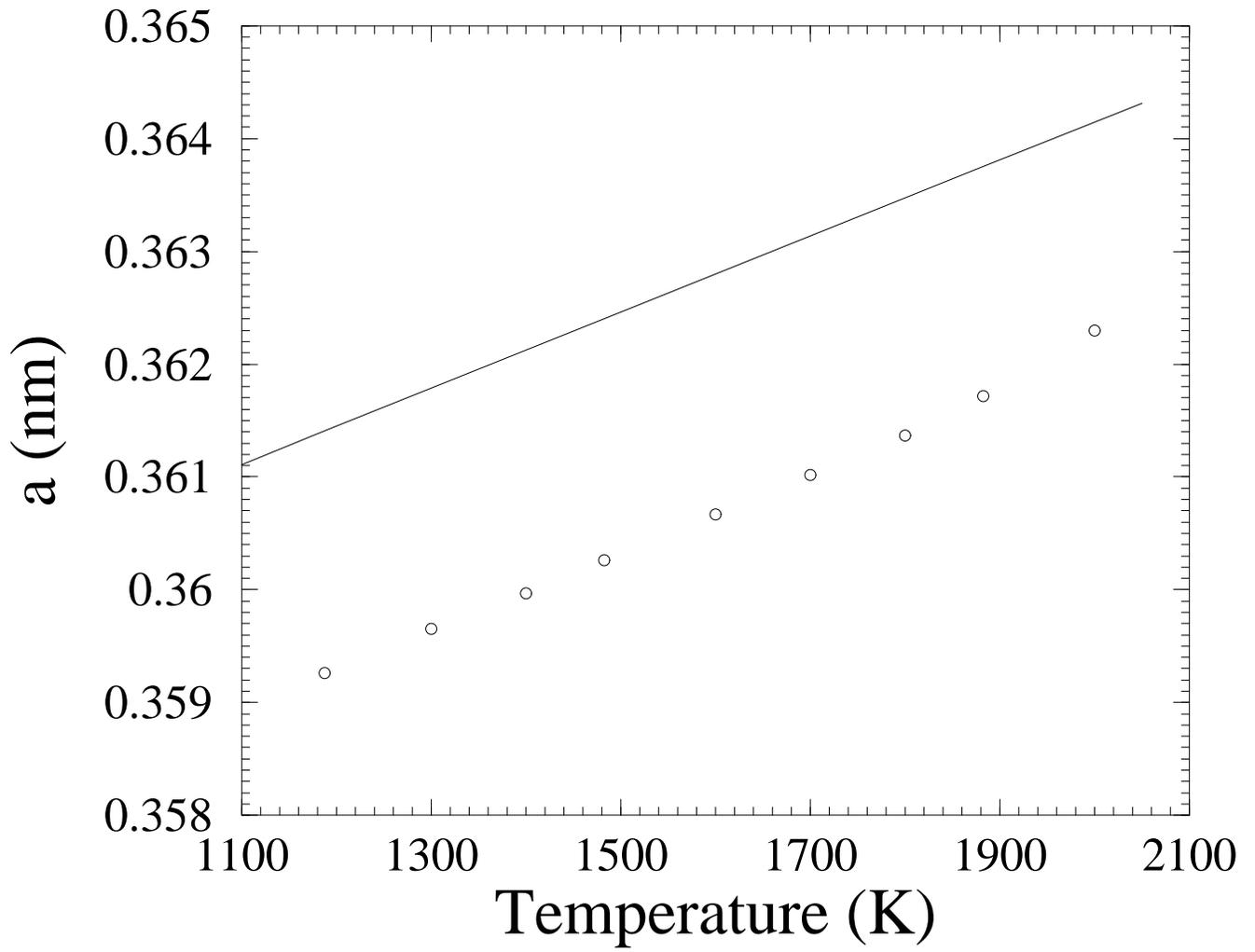}
 \caption{Lattice parameter of bcc-Zr at zero pressure $vs.$
temperature. The solid line is the experimental result and the circles are the
 computed values. The statistical error is denoted by the size of the circles.}
\label{fa(T)} 
\end{figure}

In Fig. \ref{fa(T)} we show  the computed lattice parameter vs. temperature.
Although at temperatures above 1200K the computed lattice parameter is about
 0.6\% smaller than the experimental value, the thermal expansion coefficient
 (slope) is reproduced to great accuracy $\beta = 1/V \,\,
(\partial V / \partial T) = 3.0 \cdot 10^{-5} {\rm K}^{-1}$  $(\beta_{exp} =
2.8 \cdot 10^{-5} {\rm K}^{-1}$ \cite{heiming}).
Notice that the linear extrapolation of the computed lattice parameter to 
 zero temperature does not match the fitted value. This is because at low
 temperature the thermal expansion given by the interatomic potential is
 strongly nonlinear. This is probably related to the fact that the bcc
 structure is unstable, although this behavior is not observed when comparing
 the high temperature experimental results \cite{heiming} with the zero
 temperature ab initio calculations (see Table \ref{tabIII}).

\begin{figure}
 \psboxto(0.99\textwidth;0cm){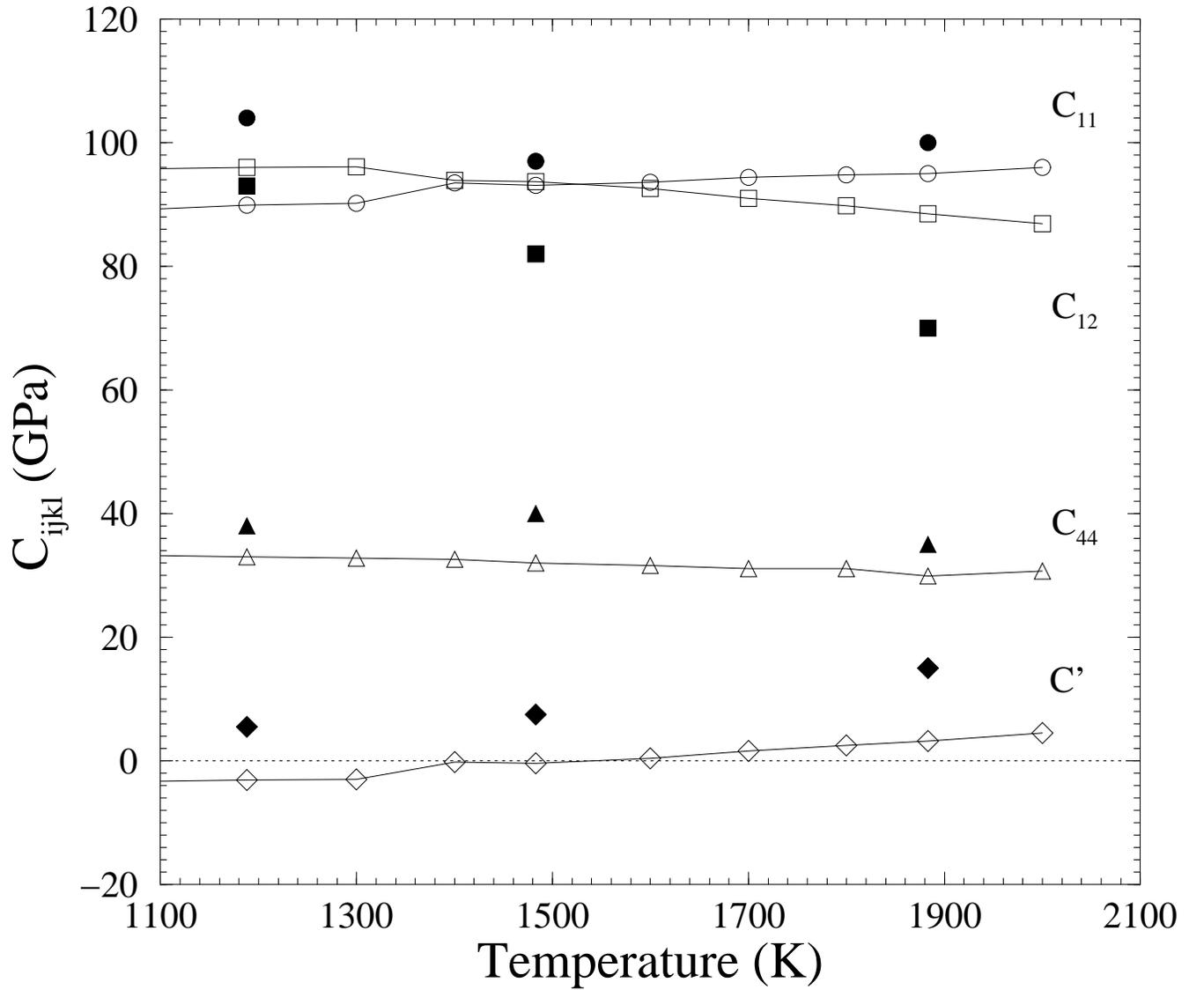}
 \caption{Isothermal second-order elastic constants of bcc-Zr
$vs.$ temperature.  Empty symbols are the computed values, and filled symbols
 are the experimental results.  
The horizontal dashed line emphasizes the change of the sign of the elastic
constant $C^\prime$ and the solid lines are guides to the eye.  
The elastic constants are $C_{11}$ (circles), $C_{12}$ (squares), $C_{44}$
(triangles) and $C'$ (diamonds). The statistical error of the computed values
is smaller than the size of the symbols}
\label{fcijkl}
\end{figure}

In Fig. \ref{fcijkl} we show the temperature dependence of the relevant 
elastic constants obtained from Monte Carlo simulations.
The main success of the present interatomic potential is that it renders a
 positive $C'$ at high temperature. Moreover, the value predicted for the whole
set of elastic constants, $C_{11}$, $C_{12}$, and  $C_{44}$ is rather accurate. The main failure is that it is unable to reproduce the large value of $C'$
observed experimentally. This failure comes mainly from the values of the
 elastic constant $C_{12}$ which experiments have shown to decrease strongly
 with temperature.
This marked decrease of $C_{12}$, together with the nearly constant
 behaviour of $C_{44}$,
 means that the Cauchy pressure decreases with temperature. We were unable
to reproduce such behaviour. In several tests during the parametrization of the
 interatomic potential we always found a Cauchy pressure nearly independent on
 temperature.

In Table \ref{tabVI} we show separately the different contributions to the
 elastic constants: the Born term, the fluctuation term, and the kinetic term.
 We get the rather unusual result that the fluctuation term of $C_{12}$ is
 nearly zero. This means that the strains in different
 directions are uncorrelated in the simulations

\begin{equation}
\left \langle \frac{\partial E_{coh}}{\partial \epsilon_{xx}}
\frac{\partial E_{coh}}{\partial \epsilon_{yy}} \right \rangle \simeq
\left \langle \frac{\partial E_{coh}}{\partial \epsilon_{xx}} \right \rangle
\left \langle \frac{\partial E_{coh}}{\partial \epsilon_{yy}} \right \rangle.
\end{equation}

Since the contribution of the fluctuation term to $C_{12}$ is usually negative
and the interatomic potential is unable to reproduce the low value
of $C_{12}$ observed experimentally, we conclude that this lack of
correlation given by the interatomic potential is possibly unphysical.

\section{Discussion and Conclusions}

 In the present paper we have developed a TB interatomic potential suitable
 for the study of the vibrational properties of bcc Zr. The interatomic
 potential has been fitted to the $T = 0$K properties of Zr in both the hcp
and the bcc structures. Although among the vibrational properties, only the
 elastic constants are used in the fitting procedure, the TB potential shows a
  remarkable capacity of predicting the $T = 0$K phonon frequencies of the
bcc structure along the high symmetry directions studied. As regards the
 high-temperature elastic constants, the general trends are reproduced, especially
 the stability of the bcc structure with respect to the shear associated
 with the elastic constant $C^\prime$. Nevertheless, the interatomic potential
 is unable to reproduce the temperature decrease of the Cauchy pressure.

 The reliability of the experimental values of the high-temperature 
 elastic constants should, however, be questioned. The elastic constants
 cannot be obtained from measurements of the velocity of acoustic waves
 in the material because the temperature at which
 the bcc phase is stable is too high. Heiming et al. \cite{heiming} therefore
 derived the elastic constants from the force constants obtained
 from a fit to the phonon dispersion relations. 
  In order to keep the number of force constants reasonably
 small, in the fitting procedure they impose that the range is up to the fifth
 neighbour shell, which is rather short. On the other hand, the 
elastic constants obtained depend critically on the frequencies of
the phonons close to the $\Gamma$ point of the Brioullin zone, and thus have
large error bars. We have tried to derive the elastic
 constants from the phonon frequencies of Heiming et al. \cite{heiming}, but we
 only were able to reproduce 
 the elastic constant $C^\prime$ to any accuracy. The values of all the other elastic
 constants strongly depend on the phonons considered and the method of
 fitting.

 In spite of  the remarkable success of the interatomic potential in reproducing
 the $T = 0$K phonon frequencies, we should mention the difficulties we have
 found during the fitting procedure, and discuss which features of the TB
 potential are expected to correctly reproduce the physics of the material
and which are not.

 The first point concerns the range of the hopping integrals, which in fact is
too small. In the hcp lattice they should fall between the second and third nearest
neighbours, at least, and only the nearest neighbours are taken into account.
 This leads to a bond energy contribution in the hcp lattice that is smaller than in
the bcc lattice, which is just the opposite to the expected result. This fact is
 compensated by the pair potential contribution to give the correct energy
difference between both structures, but it is still clearly reflected in the
 unrelaxed vacancy formation energies.

The reason for choosing such a low value for the cutoff distance of the
hopping integrals is computational convenience. The range of the
hopping integrals in the bcc lattice is up to second nearest neighbours 
(14 atoms involved). This means that in the perfect lattice at $T = 0$K
 the number of neighbours involved in the computation of the moments is 64
 (up to the sixth coordination shell). Nevertheless, at $T = 2000$K the
atoms are far from their equilibrium positions and this rapidly increases the
 number of
 neighbours involved in the computation of the moments up to $\simeq 110$.
In order to correctly compute the moments we must therefore take into account
 the neighbours up to the 13th coordination shell (258 atoms).
 An increase in the cutoff  distance of the hopping integrals involves an
increase of the number of neighbours to be taken into account in the
 computation of the moments, and thus, we decided to choose the lowest possible
 value which allowed us to obtain physically reasonable results.

 The second point concerns the pair potential contribution. During the fitting
  procedure we observed that the capacity of the interatomic potential to
 simultaneously reproduce the properties of both the hcp and the bcc structure
 is strongly dependent on the range of the pair potential. In other words, if we take
 a different range to that used in this paper, the results are rather
worse. This is indicative that the geometry of the different coordination shells
 has an important effect on the elastic constants. Moreover, although at the
 cutoff distance the pair potential and its first
derivative are continuous, the decay to zero is still sharp, and the
 contribution to the elastic constants by the last coordination shell
 is unphysically high.

 Finally, we should mention that the $s$-$d$ hybridization, which is not
 explicitly included in the TB potential, has an important  contribution to the
 cohesive energy (about -2eV \cite{gelatt}). We have considered only $d$ atomic
 orbitals in the basis set in order to minimize computation time and the
 complexity of the TB potential. 

 The problems encountered when trying to use the physical values for the
 quantities $N_d$, $W$, and $q$ have already been discussed. Nevertheless, the
treatment of these quantities as fitting parameters gives enough flexibility
 to the interatomic potential to reproduce to reasonable accuracy all the
 magnitudes described in the paper.

 A significant improvement in the behaviour of the elastic constants
 requires a better determination of the Fermi energy, together with a
more detailed description of the DOS, especially around this point.
 Nevertheless, the
inclusion of high order moments into the interatomic potential is
 computationally very expensive.

\acknowledgements We acknowledge financial support from the Comisi\'on
Interministerial de Ciencia y Tecnolog\'{\i}a (CICyT, project number 
MAT98-0315)
and supercomputing support from Fundaci\'o Catalana per a la Recerca (FCR) 
and
Centre de Supercomputaci\'o de Catalunya (CESCA).  M.P.  also acknowledges
financial support from the Comissionat per a Universitats i Recerca 
(Generalitat
de Catalunya).


\newpage

\begin{table}
\caption{Cauchy pressures of hcp Zr
  for different values of $N_d$. The parameter $q$ and the band width $W$ are
  also shown.} 
\begin{tabular}{|ccccc|}
$N_d$ & q ($a_{bcc}^{-1}$) & W (eV) & $(C_{12} - C_{66})_{hcp}^{homog.}$ (GPa) & $(C_{13} - C_{44})_{hcp}$ (GPa) \\ \hline
 1.35 & 3.730 & 13.8 & 1 & 28 \\
 1.40 & 3.680 & 13.4 & 5 & 30 \\
 1.45 & 3.634 & 13.0 & 9 & 32 \\
 1.50 & 3.592 & 12.6 & 13 & 34 \\
 1.60 & 3.498 & 12.0 & 21 & 37 \\
 1.80 & 3.342 & 10.9 & 34 & 44 \\
 2.00 & 3.201 & 10.1 & 44 & 49 \\
 2.20 & 3.099 & 9.2 & 53  & 53 \\
 2.40 & 3.012 & 8.9 & 57  & 55 \\
 2.60 & 2.953 & 8.5 & 62  & 57 
\end{tabular}
\label{tabI}
\end{table}

\begin{table} \caption{Parameters of the interatomic potential and their 
 physical values from ab initio calculations.} 

\begin{tabular}{|ccc|} 
 Parameter &  Interatomic potential  &   Ab initio \\  \hline
  $N_d$  &  1.45             &  2.536 \cite{bassani}         \\
  $q$ ($a_{bcc}^{-1}$)   &  3.634            &  $\simeq 4.36$ \cite{apettifor,pettiforllibre} \\
  $W$ (eV)   &  13.0             &  7.8 \cite{apettifor,pettiforllibre}           \\
  $r_1$ ($a_{bcc}$) &  0.92             &                 \\
  $r_2$ ($a_{bcc}$) &  1.24             &                 \\
  $A$  (eV)  &  640.476022       &                 \\
  $p_1$ ($a_{bcc}^{-1}$)  &  8.45462323       &                 \\
  $\xi$ (dimensionless) &  $-1.80733296E-05$  &                 \\
  $p_2$ ($a_{bcc}^{-1}$)  &  $-0.961306397$     &                 \\
  $\tilde{r}_1$ ($a_{bcc}$) &  1.20     &                 \\
  $\tilde{r}_2$ ($a_{bcc}$) &  2.211    &                 
\end{tabular}
\label{tabII}
\end{table}

\begin{table} \caption{Properties of bcc-Zr used for the fit.}

\begin{tabular}{|ccc|} 
  &  Interatomic potential  &   Experiment / Ab initio \\  \hline
  $a$ (nm)   &   0.3574          &  0.3574 \cite{heiming}, 0.3580 \cite{pviena}        \\
  $E_{coh}$ (eV) &  $-6.15$         &  $\simeq -6.15$ \cite{willaime91} \\
  $C_{11}$  (GPa) & 81.7  &  81.7 \cite{pviena} \\
  $C_{12}$  (GPa) &  93.4  &  93.4 \cite{pviena} \\
  $C_{44}$  (GPa) &  36   &  36 \cite{pviena} \\
  $C'$  (GPa)    &  $-5.8$   & $-5.8$  \cite{pviena} \\
  $(E_V^f)^{{\rm unrelx}}$  (eV) &  2.32  &  2.30 \cite{lebacq} \\
 \end{tabular} 
\label{tabIII}
\end{table}

\begin{table} \caption{Properties of hcp-Zr used for the fit.}

\begin{tabular}{|ccc|} 
  &  Interatomic potential &  Experiment / Ab initio \\  \hline
  $a$ (nm)     &   0.3196     & 0.3229  \cite{goldak} \\
  $c/a$    &   1.6284     & 1.592 \cite{goldak}  \\
  $(E_{coh})_{\rm bcc}-(E_{coh})_{\rm hcp}$ (eV) & 0.044  &  0.04 \cite{willaime91,moroni},  0.09 \cite{pviena} \\
  $C_{11}^{\rm homog.}$ (GPa) &  161.8  &  162  \cite{fisher,pviena} \\
  $C_{12}^{\rm homog.}$ (GPa) &  60.1  &  60  \cite{fisher,pviena} \\
  $C_{13}$ (GPa) &  68.2  &  64.6 \cite{fisher} \\
  $C_{44}$ (GPa) &  36.6  &  36.3  \cite{fisher} \\
  $C_{33}$ (GPa) &  179.5  &  172.5 \cite{fisher} \\
  $C_{66}^{\rm homog.}$ (GPa) &  50.8  &  51  \cite{fisher,pviena} \\
  $(E_V^f)^{{\rm unrelx}}$ (eV) &  2.44  &  2.07 \cite{lebacq}
 \end{tabular} 
\label{tabIV}
\end{table}

\begin{table} \caption{Force constants of bcc-Zr obtained from the tight
binding interatomic potential (in $10^{-3}$ N/m).}

\begin{tabular}{|rcrrrrrr|} 
 shell & coord.$\times$2 & xx &  yy    &   zz     &    yz     &    xz     &    xy \\ \hline
 1 & 111 & $-5495.93$ & $-5495.93$ & $-5495.93$ & $-13410.85$ & $-13410.85$ & $-13410.85$ \\
 2 & 200 & 4320.75  & $-6704.46$ & $-6704.46$ &      0    &     0     &       0  \\
 3 & 220 & $-454.85$  &  $-454.85$ & $-620.49$  &      0    &     0     &  $-932.34$ \\
 4 & 311 & $-2418.67$ &   $-95.44$ &  $-95.44$  &   $-410.94$ &  $-437.91$  &  $-437.91$ \\
 5 & 222 &   592.90 &  592.90  &  592.90  &    306.61 &    306.61 &   306.61 \\
 6 & 400 &   584.12 &  $-42.25$  &  $-42.25$  &     0     &       0   &      0   \\
 7 & 331 &   957.30 &  957.30  &  191.96  &   346.44  &   346.44  &   934.45 \\
 8 & 420 &   $-73.56$ &   21.17  &   14.13  &      0    &      0    &   $-55.33$ \\
 9 & 422 &  $-122.99$ &    7.54  &    7.54  &    $-7.77$  &   $-70.19$  &   $-70.19$ \\
 10& 333 &   $-98.35$ &  $-98.35$  &   $-98.35$ &  $-107.07$  &  $-107.07$  &  $-107.07$ \\
 10& 511 &   $-82.20$ &   10.22  &    10.22 &    $-3.17$  &   $-14.81$  &   $-14.81$ \\
 11& 440 &   $-34.35$ &  $-34.35$  &   19.05  &       0   &       0   &   $-44.28$ \\
 12& 531 &   $-29.68$ &  $-10.22$  &    3.53  &    $-3.35$  &    $-8.72$  &   $-23.40$ \\
 13& 442 &   $-14.60$ &  $-14.60$  &   $-3.05$  &    $-8.83$  &    $-8.83$  &   $-17.23$ \\
 13& 600 &   $-64.36$ &    6.10  &    6.10  &        0  &       0   &       0  \\
 14& 620 &   $-27.97$ &   $-3.05$  &    3.05  &        0  &       0   &   $-12.88$ \\
 15& 533 &    $-8.94$ &   $-3.53$  &   $-3.53$  &    $-3.53$  &  $-6.23$    &    $-6.23$ \\
 16& 622 &   $-13.71$ &   $-1.52$  &   $-1.52$  &    $-1.52$  &  $-6.30$    &    $-6.30$ \\
 17& 444 &   $-6.47$  &   $-6.47$  &   $-6.47$  &    $-6.47$  &  $-6.47$    &    $-6.47$ \\
 18& 551 &      0   &     0    &     0    &      0    &    0      &     0    \\
 18& 711 &  $-12.22$  &     0    &     0    &      0    &   $-2.66$   &   $-2.66$  \\
 19& 640 &     0    &     0    &     0    &      0    &    0      &     0    \\
 20& 642 &     0    &     0    &     0    &      0    &    0      &     0    \\
 21& 553 &     0    &     0    &     0    &      0    &    0      &     0    \\
 21& 731 &     0    &     0    &     0    &      0    &    0      &     0    \\
 22& 800 &   $-4.67$  &     0    &     0    &      0    &    0      &     0
 \end{tabular} 
\label{tabV}
\end{table}

\begin{table} \caption{Born, fluctuation, and kinetic  contributions to
the second-order isothermal elastic constants (in GPa)
 obtained from the Monte Carlo simulations in the canonical ensemble at
 different temperatures (in K) and zero pressure.}

\begin{tabular}{|cccccccccc|} 
 T & $C_{11}^{Born}$ & $C_{12}^{Born}$ & $C_{44}^{Born}$ &
 $C_{11}^{fluct.}$ & $C_{12}^{fluct.}$ & $C_{44}^{fluct.}$ &
 $C_{11}^{kin.}$ & $C_{12}^{kin.}$ & $C_{44}^{kin.}$ \\ \hline
  1188 & 126.8  &  92.6  &  48.6  & $-38.3$  &  3.4  &  $-16.3$  & 1.4 & 0 & 0.7 \\
  1300 & 128.2  &  92.4  &  48.4  & $-39.5$  &  3.7  &  $-16.4$  & 1.5 & 0 & 0.8 \\
  1400 & 129.8  &  92.2  &  48.2  & $-37.9$  &  1.7  &  $-16.5$  & 1.7 & 0 & 0.8 \\
  1483 & 131.0  &  91.8  &  48.0  & $-39.7$  &  1.9  &  $-16.8$  & 1.7 & 0 & 0.9 \\
  1600 & 131.7  &  91.7  &  47.7  & $-40.0$  &  0.9  &  $-17.0$  & 1.9 & 0 & 0.9 \\
  1700 & 132.5  &  91.5  &  47.5  & $-40.1$  & -0.4  &  $-17.3$  & 2.0 & 0 & 1.0 \\
  1800 & 133.1  &  91.1  &  47.2  & $-40.4$  & -1.4  &  $-17.2$  & 2.1 & 0 & 1.1 \\
  1883 & 133.5  &  90.7  &  46.8  & $-40.8$  & -2.2  &  $-17.9$  & 2.2 & 0 & 1.1 \\
  2000 & 135.0  &  90.8  &  46.9  & $-41.4$  & -3.9  &  $-17.4$  & 2.3 & 0 & 1.2

 \end{tabular} 
\label{tabVI}
\end{table}


\newpage

\end{document}